\documentclass{PoS}
\usepackage{amsmath}
\newcommand{\tr}{\operatorname{Tr}}
\usepackage{graphicx}
\usepackage{wrapfig}

\title{Effective SU(2) Polyakov Loop Theories with Heavy Quarks on the Lattice}

\ShortTitle{Effective SU(2) Polyakov Loop Theories with Heavy Quarks on the Lattice}

\author{\speaker{Philipp Scior}, \; David Scheffler, \; Dominik Smith\\%
        Theoriezentrum, Institut f\"ur Kernphysik, TU Darmstadt, 64289 Darmstadt, Germany\\
        E-mail: \email{scior@theorie.ikp.physik.tu-darmstadt.de, dscheff@theorie.ikp.physik.tu-darmstadt.de, smith@theorie.ikp.physik.tu-darmstadt.de}}

\author{Lorenz von Smekal\\
         Theoriezentrum, Institut f\"ur Kernphysik, TU Darmstadt, 64289 Darmstadt, Germany\\Institut f\"ur Theoretische Physik, Justus-Liebig-Universit\"at Gie{\ss}en, 35392 Gie{\ss}en, Germany\\
       E-mail: \email{lorenz.smekal@physik.tu-darmstadt.de}}

\abstract{We compare SU(2) Polyakov loop models with different effective actions with data from full two-color QCD simulations around and above the critical temperature. We then apply the effective theories at finite temperature and density to extract quantities like Polyakov loop correlators, effective Polyakov loop potentials and baryon density.}

\FullConference{The 32nd International Symposium on Lattice Field Theory\\
                 23-28 June, 2014\\
                 Columbia University New York, NY}

\begin{document}

\section{Introduction}
Phase structure and thermodynamics of QCD matter are important to understand for the phenomenology of heavy-ion collisions or the properties of neutron stars. We have significant insight into the QCD phase structure at vanishing chemical potential. Unfortunately, in the case of finite baryon chemical potential, lattice simulations are hampered by the fermion sign-problem in QCD. 
There has been a lot of effort to solve or circumvent this sign problem, such as the development of the complex Langevin algorithm for QCD \cite{Sexty2014,Aarts:2013nja}, for example. Other works have avoided the sign problem by switching to QCD-like theories, with gauge groups like  SU(2) \cite{Hands:2006ve,Cotter:2012mb} or 
 G$_2$ \cite{Wellegehausen:2013cya}, in order learn about the more  qualitative properties of strong interaction matter at finite density. 

In recent years we have also seen the development of effective Polyakov loop theories on the lattice in order to investigate the deconfinement transition of QCD. For pure gauge theories these Polyakov loop models lie in the same universality class as the underlying Yang-Mills theory. Explicit results have demonstrated that versions of these models are able to predict the location of the deconfinement transition of SU(3) Yang-Mills theory within less than 6\%  \cite{Langelage:2010yr}. It is possible to incorporate dynamical fermions in the effective theory and it has been shown that the sign problem at finite chemical potential can be dealt with, e.g.\ by a complex Langevin algorithm \cite{Langelage:2014vpa}. 

To check the range of applicability we can compare the simulations of our effective Polyakov loop theory to simulations of the underlying gauge theory. Therefore we investigate two-color effective Polyakov loop theories to be able to compare results of full two-color QCD simulations with the effective SU(2) Polyakov loop theory at all chemical potentials. 

One basic quantity to calculate is the effective Polyakov loop potential. It is a crucial input for effective theories in the continuum  like Polyakov--Quark-Meson or Polyakov--Nambu--Jona-Lasinio models. Usually one assumes the Polyakov loop potential to have no explicit dependence on the baryon chemical potential, with only implicit dependencies originating from sea quarks which are incorporated in a chemical potential dependence of the model parameters \cite{Schaefer:2007pw}. While this approximation is valid for small values of the chemical potential it is not quite clear if this is also true at larger chemical potentials.
Here we will show first steps towards such a calculation. We present unquenched results for the Polyakov loop potential obtained from full two-color lattice simulations and compare them to simulations of different effective Polyakov loop models. Further we will apply the effective theory to the cold and dense regime of two-color QCD where we measure the baryon density.

\section{Effective Polyakov Loop Theory}
The most general form of an effective action in terms of Polyakov loops is
\begin{equation}
	S_\text{eff}= \sum_{ij}L_i K^{(2)}(i,j)L_j + \sum_{ijkl} L_i L_j K^{(4)}(i,j,k,l) L_k L_l + \dots + \sum_i h^{(1)}(i) L_i + \dots \; ,
\end{equation}
with $L_x$ being the Polyakov loop at position $x$. 
	\begin{figure}[t]
	\vspace*{-0.4cm}
\begin{minipage}{0.49\textwidth}
	\centering
	\includegraphics[width=0.8\textwidth]{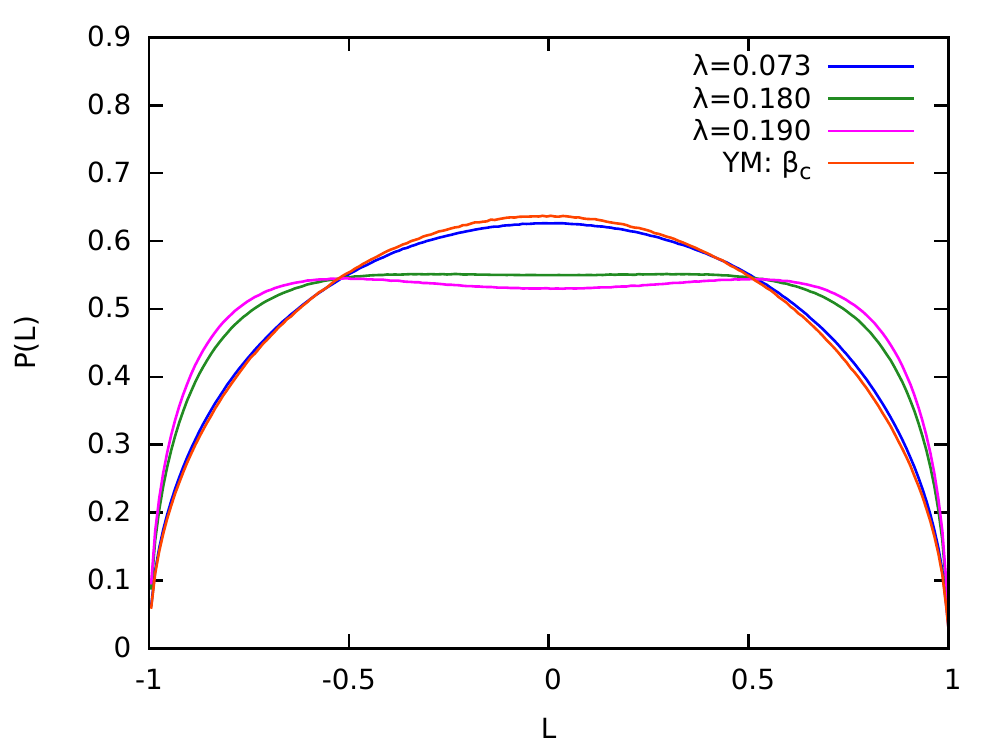}
	\caption{\label{fig_plain} Distributions of the SU(2) Polyakov loop from the simple effective model at subcritical couplings  $\lambda < \lambda_{c}=0.196$. }

\end{minipage} \hspace*{0.1cm}
\begin{minipage}{0.49\textwidth}
	\centering
	\includegraphics[width=0.8\textwidth]{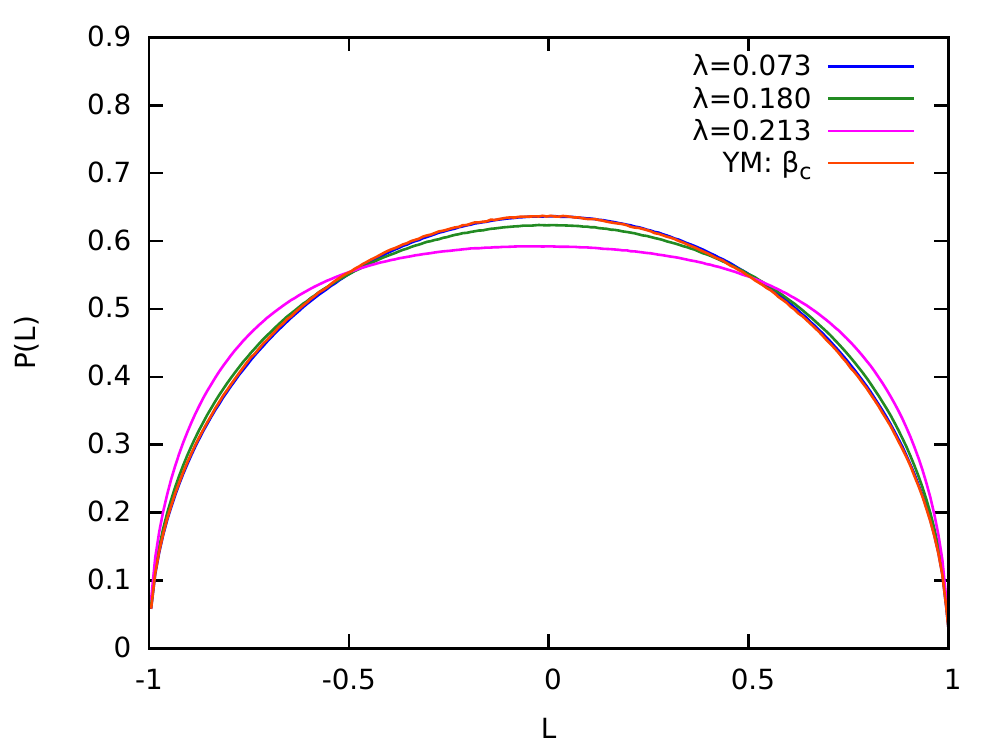}
	\caption{\label{fig_resum} Distributions of the SU(2) Polyakov loop from the resummed effective model at subcritical couplings $\lambda < \lambda_c=0.2143$. }
\end{minipage}
\end{figure}
The goal is now to find the effective kernels $K^{(n)}$ and couplings $h^{(n)}$ of the effective theory in terms of the parameters of the underlying theory. This can be done by non-perturbative methods like inverse Monte-Carlo \cite{Heinzl:2005xv} or the relative weights method \cite{Greensite:2013yd}, or one can calculate the kernels and couplings analytically in a combined strong-coupling and hopping expansion.
To compare these methods we will match results from SU(2) Yang-Mills theory to results both from the simplest ansatz for an effective action as well as the leading order result of the effective theory derived by a strong-coupling expansion. The action for the simplest Ansatz is given by
\begin{equation}
	S_\text{eff}= - \lambda \sum_{ij} L_i L_j \;, \label{plain_action}
\end{equation}
describing a nearest-neighbor interaction between Polyakov loops. The effective coupling $\lambda$ in terms of the original lattice coupling $\beta$ and time-like extension of the lattice $N_	t$ can be obtained by inverse Monte-Carlo methods. The leading order effective action from the strong-coupling approach is given by
\begin{equation}
	S_\text{eff}= - \sum_{ij} \log (1+\lambda L_i L_j) \;. \label{resum_action}
\end{equation}
Here one resums generalized Polyakov loops, winding several times around the lattice to produce the logarithm. The effective coupling in terms of the original lattice parameters is given by
\begin{equation}
	\lambda(u,N_{t}\geq5)=u^{N_t}\exp\left[N_{t}\left(4u^4-4u^6+\frac{140}{3}u^8-\frac{36044}{405}u^{10} + \mathcal{O}(u^{12})\right)\right]\;, \label{coup}
\end{equation}
\begin{wrapfigure}{r}{0.4\textwidth}
\vspace*{-.2cm}
\includegraphics[width=0.4\textwidth]{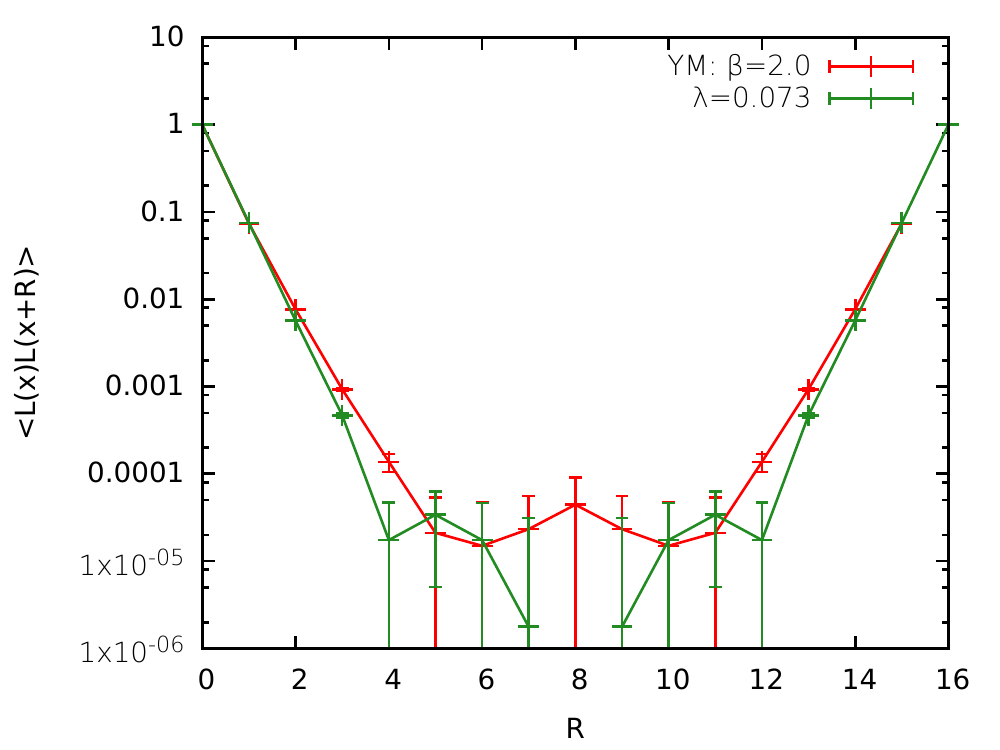}
\caption{\label{poly_corr} Polyakov-loop correlator of the effective model (\protect\ref{resum_action}) compared to the pure gauge theory at $\beta=2.0$ on a $16^3\times 4$ lattice.}
\vspace*{-.2cm}
\end{wrapfigure}
where $u(\beta)$ is the ratio of the first two modified Bessel functions $u(\beta)={I_2(\beta)}/{I_1(\beta)}$, as usual. Terms involving more Polyakov loops or longer range interactions are of higher order in $\beta$ and are therefore suppressed in the strong coupling limit. To compare the effective theories with pure gauge theory results we compare measured Polyakov loop distributions which we found to be much more sensitive to parameter changes than e.g.\ the expectation values of the Polyakov loop. In the pure gauge theory the distribution remains symmetric and unmodified throughout the symmetric phase.
	\begin{figure}[t]
	\vspace*{-0.4cm}
\begin{minipage}{0.49\textwidth}
	\includegraphics[width=0.86\textwidth]{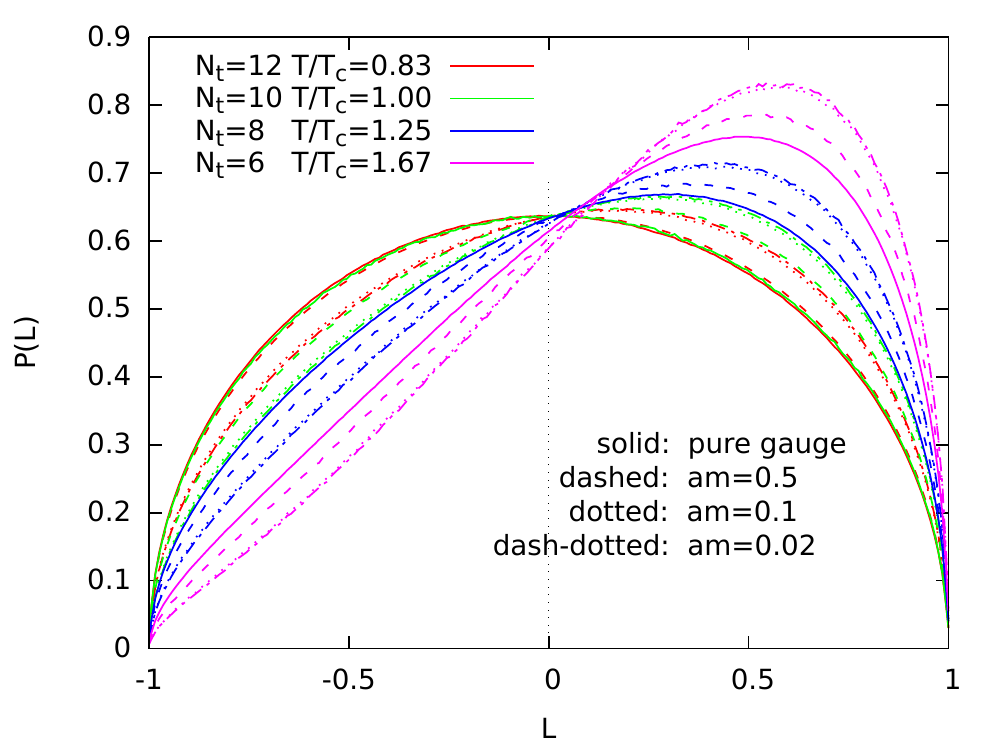}
	\caption{\label{dist_unq} Polyakov-loop distributions from simulations of two-color QCD with two flavors of staggered quarks of various masses at $\beta=2.577856$.}
\end{minipage} \hspace*{0.1cm}
\begin{minipage}{0.49\textwidth}
	\includegraphics[width=0.86\textwidth]{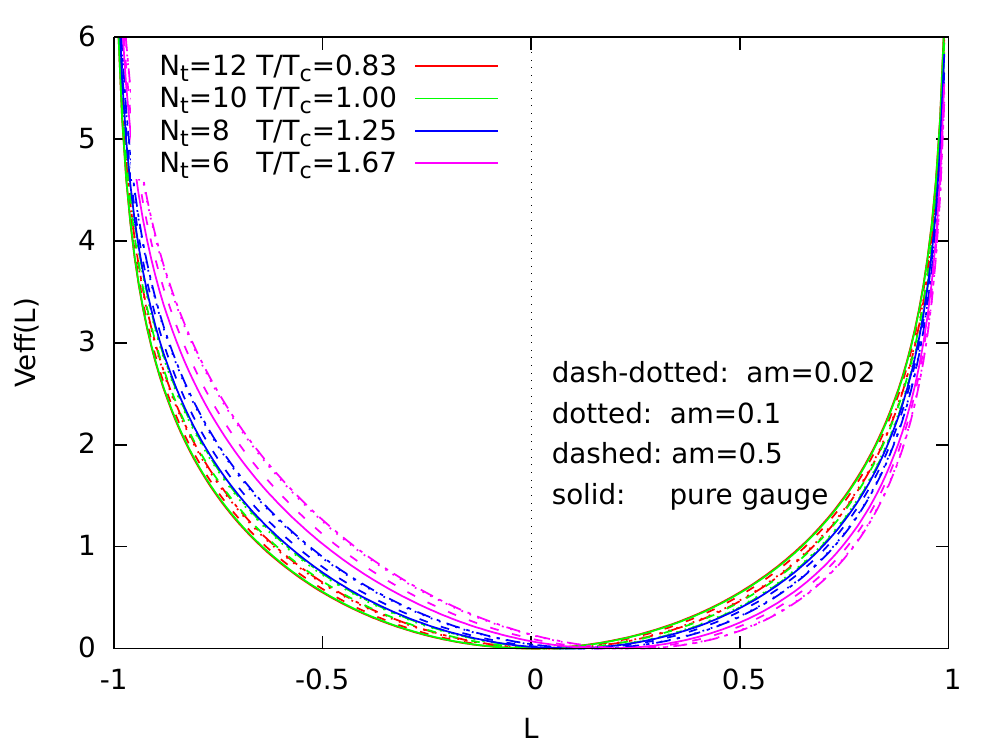}
	\caption{\label{pot_unq} Effective Polyakov-loop potentials
 for two-color QCD with two flavors of staggered quarks 
   from the Polyakov-loop distributions in Fig.~\protect\ref{dist_unq}.}
\end{minipage}
\end{figure}
At couplings above the critical coupling $\beta_c$ center symmetry is spontaneously broken and the Polyakov loop distributions get skewed, resulting in a finite value of $\langle L\rangle$. Figures \ref{fig_plain} and \ref{fig_resum} show the results of the simulations at subcritical values of $\lambda$ for both effective models compared to results from pure SU(2) gauge theory simulations at the critical coupling $\beta_c$ for $N_ t=10$. In the strong coupling regime or respectively at small effective coupling $\lambda$, the distributions match the `exact' distribution from the SU(2) gauge theory simulations very well. At larger couplings but still in the symmetric phase, the effective theory distributions get deformed, however. This effect is strongest the in simple effective model of Eq.~\eqref{plain_action} whose distribution shows strong deformations even developing  a double peak structure close to $\lambda_c$. The distributions from the resummed model, Eq.~\eqref{resum_action}, also get deformed but their overall shape remains qualitatively almost unchanged, and the deviations from the full gauge theory result are generally  much smaller. We therefore conclude that the resummed model is better suited for couplings close to the critical coupling. The Polyakov-loop correlator of this model in the strong coupling regime is compared with the gauge theory result in Fig.~\ref{poly_corr}.

\section{Effective Polyakov Loop Potential}
We can use the Polyakov loop distribution $P(L)$ to calculate the on-site effective Polyakov-loop potential. First, we calculate the constrained effective potential $V_0$ and then obtain  the effective Polyakov loop potential $V_\text{eff}$ via Legendre transformation:
\begin{align}
V_0(L)&= - \log P(L) \; , \notag \\
 W(h) &= \log \int d L' \exp(-V_0(L')+hL') \; , \\
 V_\text{eff}(L) & = \text{sup}_h \left( L h - W(h) \right) \notag \; .
 \end{align}
Note that the minimum of $V_\text{eff}$ is correctly located at the expectation value of the Polyakov loop $\langle L\rangle$. Figures \ref{dist_unq} and \ref{pot_unq} show Polyakov loop distributions and the corresponding effective potentials from full two-color QCD simulations with two flavors of staggered quarks at $\beta=2.577856$ and different masses. The relative temperature scale is taken from \cite{Smith:2013msa}. There is a clear trend, the distributions get skewed to the right resulting in a finite $\langle L\rangle$. This effect gets stronger for higher temperatures and smaller quark masses. The effect on the effective potential is similar but not quite as drastic: The minimum of the potential moves to larger $L$ with higher temperature and lower quark mass.

 Since we now want to analyze the unquenched effective Polyakov loop potential with our effective theory we have to include dynamical fermions into the theory. The structure of the fermion action can be derived from a combined strong coupling and hopping expansion analogous to \cite{Langelage:2014vpa}. As for the pure gauge theory on can resum generalized Polyakov loops to obtain,
	\begin{figure}[t]
	\vspace*{-0.4cm}
\begin{minipage}{0.49\textwidth}
	\includegraphics[width=0.8\textwidth]{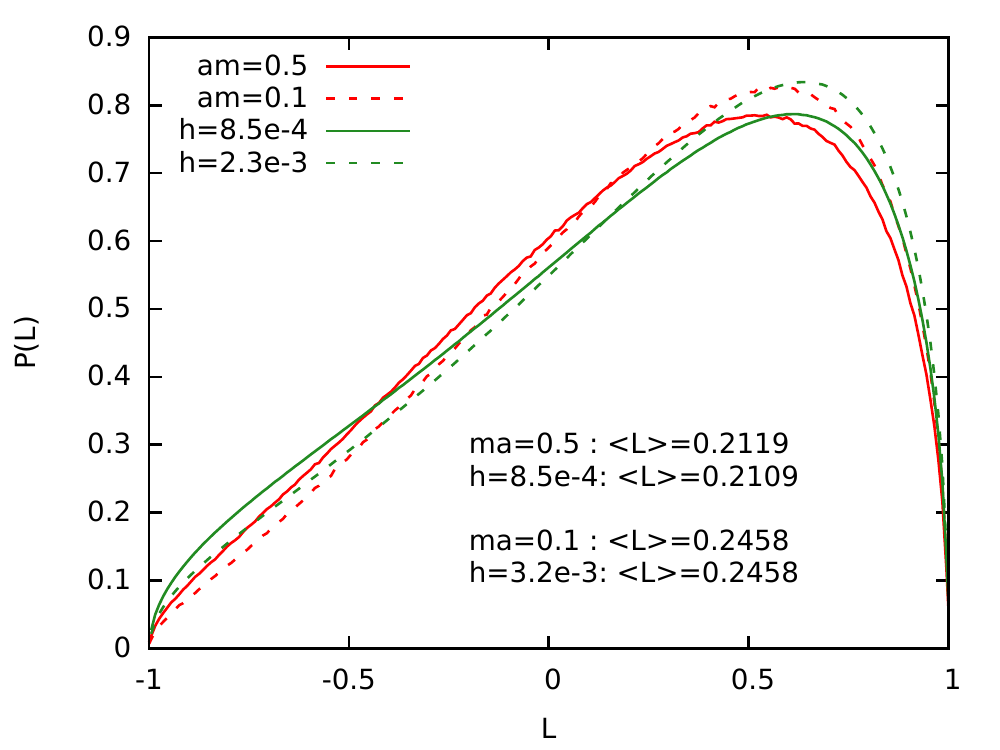}
	\caption{\label{dist_eff} Polyakov-loop distributions from the effective theory compared to two-color QCD simulations at $T=1.67 \, T_c$ with $am=0.1$ and $am=0.5$.} 
\end{minipage} \hspace*{0.1cm}
\begin{minipage}{0.49\textwidth}
	\includegraphics[width=0.8\textwidth]{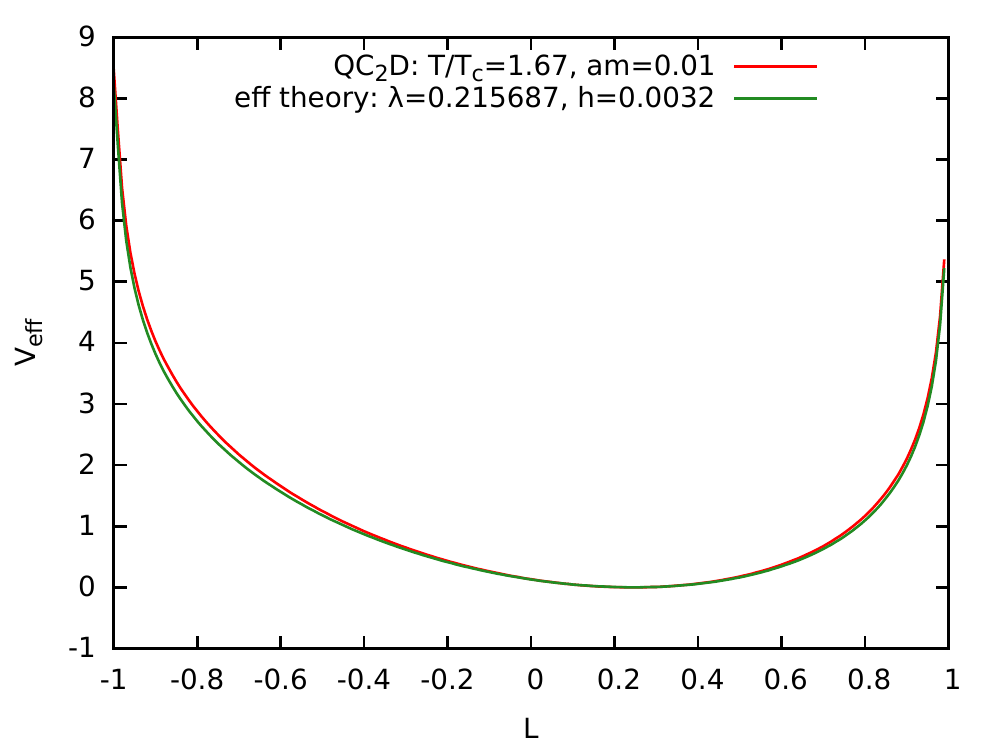}
	\caption{\label{pot_eff} Effective Polyakov loop potentials from the unquenched effective model and two-color QCD simulations at $T=1.67\, T_c$  and $am=0.1$}
\end{minipage}
\end{figure}
\hspace{0.07cm} 
\begin{equation}
	S_\text{ferm}= - 4 N_f \sum_i \log (1 + h L_i + h^2) \; , \label{ferm_action}
\end{equation}
and the leading order hopping expansion result for $h$ reads
\begin{equation}
	h(u,\kappa,N_t	)= (2 \kappa e^{a \mu})^{N_t} + \mathcal{O}(\kappa^2 u) \; . \label{hop}
\end{equation}
In the range where the hopping expansion is valid one can check that the resummation is not as important as in the gauge action because $h$ is relatively small. This changes for smaller quark masses or finite chemical potential $\mu$, however, where $h$ increases and resummation becomes important.

We choose \eqref{resum_action} together with \eqref{ferm_action} as the action for our Polyakov-loop effective theory. However, to compare our results to the full two-color results from above we have to apply the theory outside the regime where the strong coupling and hopping expansions are valid. Therefore, relations \eqref{coup} and \eqref{hop} for the effective couplings are no longer valid. Our strategy for comparing our full two-color simulations with the ones from the effective theory is a two step procedure: First, we compare distributions from pure gauge simulations and the effective theory without quarks to match distributions with the same expectation value of the Polyakov loop,
\begin{equation}
	\langle L\rangle  = \int_{-1}^1 dP(L) \; L \; ,
\end{equation}
in order to find the effective gauge coupling $\lambda$ associated with a particular temperature $T$. In the second step we then match the effective theory with dynamical quarks at that $\lambda$ or $T$ for different values of $h$ to the distributions from full two-color QCD simulations with a particular quark mass $am$, again identifying distributions with the same $\langle L\rangle$. Figures \ref{dist_eff} and \ref{pot_eff} show Polyakov-loop distributions and effective potentials from the effective model compared to two-color QCD results for a temperature of $T= 1.67\, T_c$. The overall shape of the distribution is reproduced quite well, but some deviations from the full two-color QCD results remain. One can show, that these deviations originate in the gauge part of the action. We believe that this is due to neglecting the interactions between Polyakov loops at larger distances which become increasingly important with higher temperatures, above  $T_c$. 
The effective potential shows much better agreement than the distributions. It seems to be a generic effect of the Legendre transformation that distributions with the same expectation value $\langle L\rangle$ lead to quite similar effective potentials.

\section{Effective Theory for the Cold and Dense Regime}
We will now apply the effective Polyakov-loop theory in the cold and dense regime of two color QCD with heavy quarks. Here we are well inside the region where the strong coupling and hopping expansions are applicable. In fact, the effective gauge coupling $\lambda$ from \eqref{coup} is negligible even at $\beta=2.5$ in the temperature range we are interested in ($T \sim 4 - 10$ MeV). We have $\lambda(\beta=~2.5,N_t=~200)\sim 1 \cdot 10^{-15}$. We therefore end up with a completely fermionic partition function. Due to the large number of time slices, with $N_t$ between $200$ and $600$, we have to include more terms in the hopping expansion of the fermion determinant. The effective action up to order $\kappa^4$ in the hopping exansion reads
\begin{figure}[t]
	\vspace*{-0.3cm}
\begin{minipage}{0.49\textwidth}
	\vspace*{-0.4cm}
	\centering
	\includegraphics[width=0.85\textwidth]{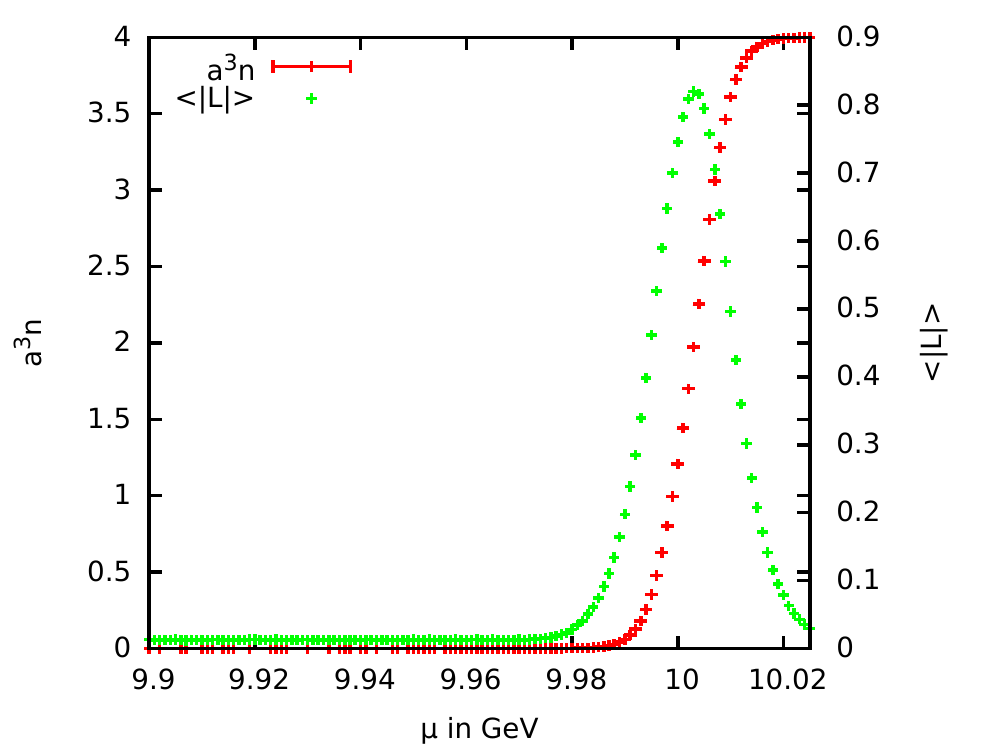}
	\caption{\label{linear} Quark density in lattice units $a^3n$ together with the Polyakov loop $\langle |L| \rangle$ as a funtion of $\mu$ in physical units.}
\end{minipage} \hspace*{0.1cm}
\begin{minipage}{0.49\textwidth}
\centering
	\includegraphics[width=0.8\textwidth]{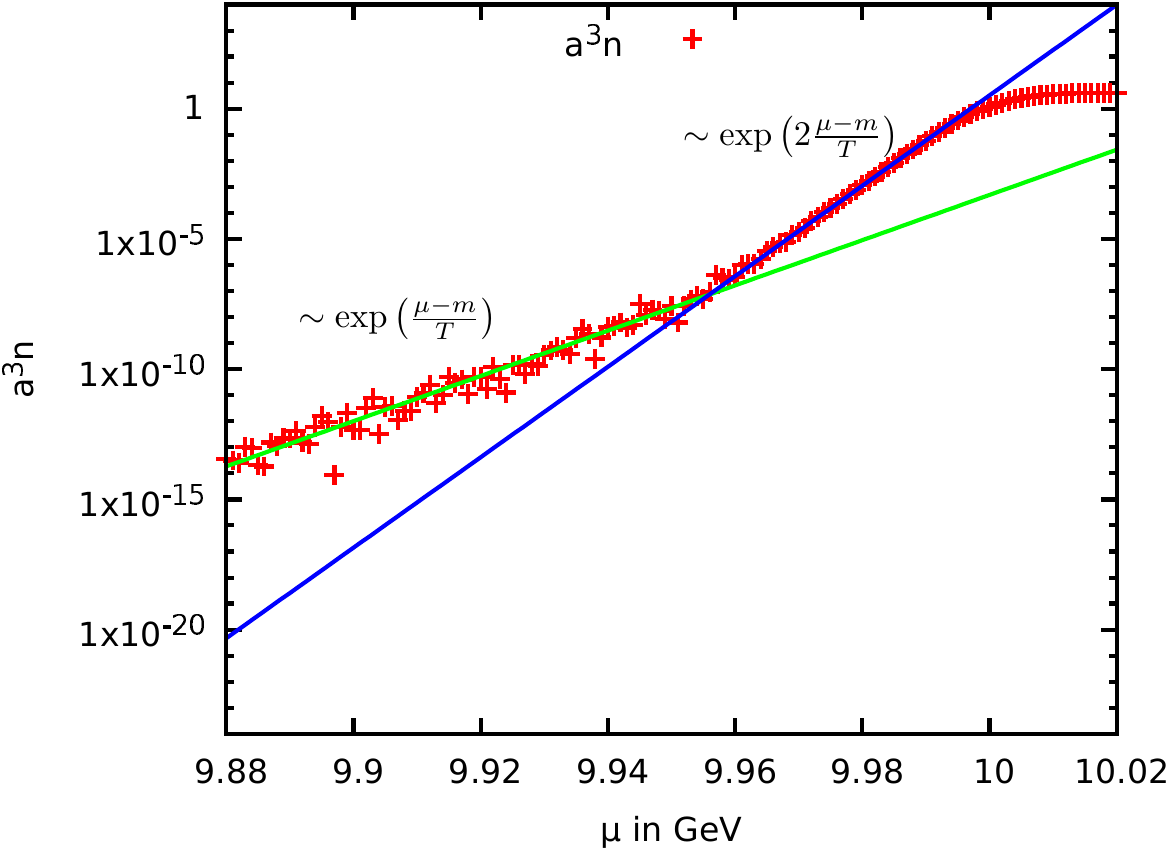}
	\caption{\label{slopes} Logarithmic plot of the density in lattice units $a^3n$. The lines indicate the two regions with different exponential increases.}
\end{minipage}
\end{figure}

{ \small
\begin{align}
&-S_\text{eff}= \sum_{\vec x} \log (1+hL_i +h^2)^2 -2 h_2 \sum_{\vec x, i} \tr \frac{h W_{\vec x}}{1 +h W_{\vec x}} \tr \frac{h W_{\vec x+i}}{1+h W_{\vec x+i}} 
+2 \frac{\kappa^4 N_t^2}{N_c^2} \sum_{\vec x, i} \tr \frac{h W_{\vec x}}{(1 +h W_{\vec x})^2} \tr \frac{h W_{\vec x+i}}{(1+h W_{\vec x+i})^2} \notag \\
&+\frac{\kappa^4 N_t^2}{N_c^2} \sum_{\vec x, i, j} \tr \frac{h W_{\vec x}}{(1 +h W_{\vec x})^2} \tr \frac{h W_{\vec x-i}}{1+h W_{\vec x-i}} \tr \frac{h W_{\vec x-j}}{1+h W_{\vec x-j}} 
+2\frac{\kappa^4 N_t^2}{N_c^2} \sum_{\vec x, i, j} \tr \frac{h W_{\vec x}}{(1 +h W_{\vec x})^2} \tr \frac{h W_{\vec x-i}}{1+h W_{\vec x-i}} \tr \frac{h W_{\vec x+j}}{1+h W_{\vec x+j}} \notag \\
&+\frac{\kappa^4 N_t^2}{N_c^2} \sum_{\vec x, i, j} \tr \frac{h W_{\vec x}}{(1 +h W_{\vec x})^2} \tr \frac{h W_{\vec x+i}}{1+h W_{\vec x+i}} \tr \frac{h W_{\vec x+j}}{1+h W_{\vec x+j}} 
+\kappa^4 N_t \sum_{x,i} \frac{h^4}{(1+h L_x+h^2)(1+h L_{x+i}+h^2)} \; , \label{eff_action} 
\end{align} } 

where $W_{\vec x}$ stands for an untraced Polyakov loop. The only leftovers from the Yang-Mills part of the original theory in the effective action are gauge corrections to the effective fermion couplings
\begin{align}
 	h &= \exp \left[ N_t \left(a \mu + \ln 2 \kappa + 6 \kappa^2 \frac{1-u^{N_t}}{1-u} \right) \right] \; ,   \quad 
 	h_2= \frac{\kappa^2 N_t}{N_c}\left[1+2 \frac{u-u^{N_t}}{1-u}+ \dots \right] \; .
 \end{align} 
In order to set a physical scale we again use the scale from \cite{Smith:2013msa} together with $\sqrt{\sigma}=440$ MeV and we calculate the diquark mass in the combined strong coupling and hopping expansion to be
\begin{equation}
 a m_d= - 2 \ln(2\kappa) - 6 \kappa^2 -24 \kappa^2 \frac{u}{1-u} + 6 \kappa^4 + \dots \; . \label{mass}
 \end{equation} 
We are now able to determine the fermion number density of the system,
\begin{equation}
	n = \frac{T}{V} \frac{\partial}{\partial \mu} \log Z \; .
\end{equation}
Figures \ref{linear} and \ref{slopes} show the results for a simulation in the cold and dense regime. The parameters for this simulation are $\beta=2.5$, $a=0.081$ fm, $\kappa=0.00802$, $m_d=20$ GeV, $N_t=484$, $T=5$ MeV. Figure \ref{linear} shows a sharp increase in the density just below $\mu=m_d/2$ in agreement with the Silver Blaze property. 
However, in our opinion this sharp increase should not be interpreted as the onset of Bose-Einstein-Condensation of diquarks, i.e.\ nuclear matter in two-color QCD, because the Polyakov-loop expectation value starts increasing even below this point indicating deconfinement. To get a better insight of what is going on, we can look at a logarithmic plot of the quark density shown in Figure \ref{slopes}. For chemical potentials $\mu<9.96$ GeV the slope is well described by a free quark gas. This comes from the small but finite Polyakov-loop expectation value at finite $T$ due to the presence of the dynamical quarks which break center symmetry explicitly. At around $\mu \approx 9.96$ GeV the density has a `kink' from where on it is dominated by two-quark states indicative of statistical confinement. Since it is described by a thermal distribution in this regime as well, it does not represent a diquark BEC but 
 a thermal diquark gas at $T\neq 0$. 
At very large chemical potentials the density saturates when every lattice site is occupied by the maximum number of quarks, $ 2 N_f \cdot N_c=4$ here, as known from the two-color QCD \cite{Hands:2006ve,Cotter:2012mb} or G$_2$-QCD \cite{Wellegehausen:2013cya} simulations.

\section*{Acknowledgments}
This work was supported by the Helmholtz International Center for FAIR within the LOEWE initiative of the State of Hesse. All results were obtained
on Nvidia GTX or Tesla graphics cards.

\end{document}